\documentclass{aa}
\input{psfig}

\def\tvi(#1,#2){\vrule height #1pt depth #2pt width 0pt}

\def\e{{\rm e}}

\def\ie{i.e. }
\def\eg{e.g. }
\def\etal{et al. }

\def\rso{r_{\rm son}}
\def\cs{c_{\rm son}}
\def\M{{\cal M}}

\def\cpsh{{\cal R}_{\rm sh}}
\def\cssh{{\cal Q}_{\rm sh}}

\def\rfr{{\bar{\cal R}}_{a}}
\def\rfsh{{\bar{\cal R}}_{\rm sh}}
\def\qfq{f_{\rm sh}}

\newlength{\largeur}
\newlength{\saut}
\def\marge#1{
\setlength{\largeur}{\columnwidth}
\addtolength{\largeur}{-#1}
\setlength{\saut}{0.5\largeur}\hspace*{\saut}} \def\picture #1 by #2
(#3){
\marge{#1} \vbox to #2{
\hrule width #1 height 0pt depth 0pt
\vfill
\special{picture #3}}}
\def\scaledpicture #1 by #2 (#3 scaled #4){{
\dimen0=#1 \dimen1=#2
\divide\dimen0 by 1000 \multiply\dimen0 by #4 \divide\dimen1 by 1000
\multiply\dimen1 by #4 \picture \dimen0 by \dimen1 (#3 scaled #4)}}

\begin{document}

\thesaurus{06
(02.01.2;
02.08.1;
02.09.1;
02.19.1;
08.02.1;
13.25.5)}

\title{Entropic-Acoustic instability in shocked accretion flows}
\author{T. Foglizzo\inst{1}\thanks{e-mail: {\tt foglizzo@cea.fr}}
\and M. Tagger\inst{1}\thanks{e-mail: {\tt tagger@cea.fr}} 
}

\offprints{T.~Foglizzo}

\institute {Service d'Astrophysique, CEA/DSM/DAPNIA (CNRS URA 2052), 
CE-Saclay, 91191 Gif-sur-Yvette, France 
}

\date{Received 15 May 2000; accepted 7 September 2000}
\titlerunning{Entropic-Acoustic instability}
\maketitle

\begin{abstract}
A new instability mechanism is described in accretion flows where the 
gas is accelerated from a stationary shock to a sonic surface. The 
instability is based on a cycle of acoustic and entropic waves in 
this subsonic region of the flow. 
When advected adiabatically inward, entropy perturbations trigger 
acoustic waves propagating outward. If a shock is present at the outer 
boundary, acoustic waves reaching the shock produce new entropy 
perturbations, thus creating an entropic-acoustic cycle between the 
shock and the sonic surface. The interplay of acoustic and entropy 
perturbations is estimated analytically using a simplified model based 
on the compact nozzle approximation. According to this model, the 
entropic-acoustic cycle is unstable if the sound speed at the sonic 
surface significantly exceeds the sound speed immediately after the shock.
The growth rate scales like the inverse of the advection time from 
the outer shock to the sonic point. The frequency of the most unstable 
perturbations is comparable to the refraction cutoff, defined as the 
frequency below which acoustic waves propagating inward are significantly 
refracted outward. This generic 
mechanism should occur in Bondi-Hoyle-Lyttleton accretion, and also 
in shocked accretion discs.


\end{abstract}

\section{Introduction}

Hydrodynamic instabilities in the process of accretion onto a compact 
star may help to understand the time variability of the emission from 
X-ray binaries. For example, the supersonic motion of a compact star 
in a uniform gas, first described by Hoyle \& Lyttleton (1939), 
Bondi \& Hoyle (1944), was later found to be unstable in numerical 
simulations (Matsuda \etal 1987, Fryxell \& Taam 1988, Matsuda \etal 
1992, Ruffert \& Arnett 1994). In such flows, the gas captured by the 
accretor is first heated and decelerated to subsonic velocities through a shock, 
and then accelerated to supersonic velocities towards the accretor.
Observationnal consequences of the instability were discussed by 
Taam \etal (1988), Livio (1992), De Kool \& Anzer (1993). 
Nevertheless, this instability is still poorly understood even in its 
simplest formulation. A recent attempt by Foglizzo \& Ruffert (1999) 
has led us to look in more detail at the interplay of acoustic 
and entropy waves in the subsonic flow between the bow shock and the 
accretor. We identify in the present paper an entropic-acoustic cycle:
entropy perturbations advected with the flow trigger acoustic waves propagating 
back to the shock, where in turn they trigger new entropy perturbations. 
The linear stability of this cycle depends on the efficiencies 
of these two processes, measuring:
\par(i) the amplitude of the pressure perturbations $\delta p^{-}$ propagating 
against the stream, triggered by the advection of an entropy wave of amplitude 
$\delta S_{1}$,
\par(ii) the amplitude of the entropy perturbation $\delta S_{2}$ 
triggered by the interaction of an acoustic wave $\delta p^{-}$ 
with a stationary shock.\\
The entropic-acoustic cycle is a priori unstable if 
$\delta S_{2}>\delta S_{1}$. As will be seen throughout this 
paper, a complete description of the entropic-acoustic instability requires 
to take into account acoustic waves $\delta p^+$ propagating with the stream, 
triggered by the interaction of $\delta p^{-}$ with the shock and refracted 
out by the flow gradients.\\
Since such entropic-acoustic cycles may be present in astrophysical 
flows other than the Bondi-Hoyle-Lyttleton accretion, the present paper 
is dedicated to a description of this generic mechanism in its most 
general formulation. Similar entropic-acoustic cycles have been studied 
extensively in the context of jet nozzles, where the gravitational 
potential plays no role. In particular, the first process (i) is well 
known among the combustion community since the works of Candel (1972) 
and Marble (1973). A clear overview of the subject can be found in Marble 
\& Candel (1977). An example of entropic-acoustic instability in 
nozzles is the ramjet 'rumble' instability studied by Abouseif, Keklak 
\& Toong (1984), where the combustion zone plays the same role as the 
shock in the second process (ii).\\
The paper is organized as follows: in Sect.~\ref{Sconvent} we analyze 
the excitation of acoustic waves by the advection of entropy
perturbations. The entropy perturbations 
produced by a sound wave reaching a shock are estimated in 
Sect.~\ref{sect3}. These two processes are combined into a global mode in 
Sect.~\ref{sect4}, where we characterize its spectrum of eigenfrequencies.
Possible astrophysical applications are discussed in 
Sect.~\ref{Sconclusion}.

\section{Acoustic energy released by the advection of entropy 
perturbations in an irrotational flow\label{Sconvent}}

\subsection{Energy released by the advection of a localized perturbation of 
entropy\label{Senerg}}

\begin{figure}
\psfig{file=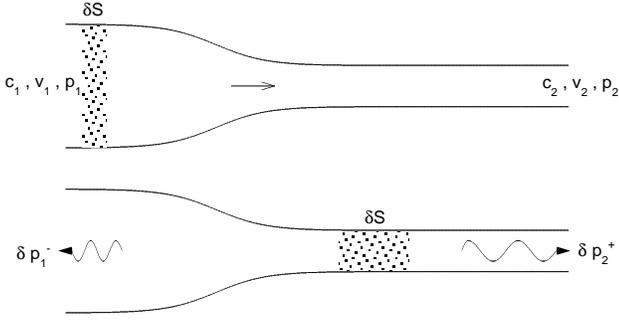,width=\columnwidth}
\caption[]{Illustration of the release of acoustic energy by the 
advection of a localized perturbation of entropy in a nozzle}
\label{figcpl}
\end{figure}
We wish to estimate the energy $\delta E_{12}$ released by the advection of 
a perturbation $\delta S$ of entropy into a gravitational potential, from 
a region I where the unperturbed velocity, sound speed and gravitational 
potential are $v_1,c_1,\Phi_1$ to a region II closer to the accretor where 
their respective values are $v_2,c_2,\Phi_2$. These quantities are related 
through the conservation of the Bernoulli constant $B$ in the unperturbed 
flow. Denoting by $\gamma$ the adiabatic index of the gas, and by $h$ 
its enthalpy,
\begin{eqnarray}
B&\equiv& {v^2\over2}+h+\Phi,\label{Bernoulli}\\
h&\equiv& {c^2\over\gamma-1}.
\end{eqnarray}  
A direct calculation of $\delta E_{12}$ is possible when the unperturbed 
flow is uniform before and after the region of convergence, so that 
the entropy perturbation has enough time to reach a pressure 
equilibrium with the surrounding gas (see Fig.~\ref{figcpl}).
Let $m+\delta m$ be the mass of the gas with a perturbed entropy 
$S+\delta S$. From the conservation of entropy,
\begin{equation}
{\delta m\over m} = \exp\left(-{\gamma-1\over\gamma}{\mu\delta S
\over{\cal R}}\right) -1.
\end{equation}
For the sake of simplicity, the ratio of the molecular weight $\mu$ to 
the gas constant ${\cal R}$ is set to $\mu/{\cal R} = 1$ throughout 
this paper, with no loss of generality.  
The total energy $\delta E_{12}$ released during the advection 
through the nozzle is calculated in Appendix~A:
\begin{equation}
\delta E_{12}=(h_2-h_1)\delta m.\label{deltae3}
\end{equation}
This calculation, which involves no linearization, indicates that the 
the release of energy during the advection of a localized entropy 
perturbation is a fundamental process associated to the variation of 
enthalpy in the flow. $\delta E_{12}$ corresponds to the sum of the 
energies of two localized pressure perturbations propagating with and 
against the stream. According to Eq.~(\ref{deltae3}), the bigger the 
increase of entropy through the flow, the bigger the release of acoustic energy 
by the advection of entropy perturbations.
In an accretion flow governed by a gravitational potential 
$\Phi\equiv GM/r$, the lengthscale of the inhomogeneity of the flow scales like 
the distance $r$ to the accretor. Consequently there is no  
region II where the flow is uniform as in Fig.~\ref{figcpl}. 
Nevertheless, the time needed by an entropy perturbation to reach a 
pressure equilibrium with the surrounding gas is of the order of the sound 
crossing time $\delta l/c$, 
where $\delta l$ is the length of the entropy perturbation along the flow. 
The distance it travels before the pressure equilibrium is 
reached is of the order of  $\M\delta l$, where $\M\equiv v/c$ 
is the mach number of the flow. The above calculation of $\delta E_{12}$ is 
thus relevant for an accretion flow in the region where the length of the 
perturbation is short enough compared to the scale of the flow inhomogeneities:
\begin{equation}
\M{\delta l\over r} \ll 1.\label{critloc}
\end{equation}
Note that Eq.~(\ref{deltae3}) does not specify the relative 
amplitudes of the pressure perturbations propagating with and against 
the stream.

\subsection{Acoustic flux released by the advection of an entropy wave}

If the entropy perturbation were an extended entropy wave, 
we expect acoustic waves to be triggered by the enthalpy gradient, 
by the same process as described in Sect.~\ref{Senerg}.
By definition, the acoustic energy of an acoustic 
wave is the part of the energy which depends (in a quadratic way) only on 
the linear approximation of this wave (\eg Lighthill 1978, Chap.~1.3). The 
flux $F^\pm$ of this acoustic energy is defined by:
\begin{equation}
F^\pm\equiv{\dot M}_0c^2{(1\pm\M)^2\over \M}
\left({\delta p^\pm\over \gamma p}\right)^2,\label{defflux}
\end{equation}
where ${\dot M}_0$ is the mass flux and ${\cal M}\equiv v/c$ the Mach number.
The index ($+$) is used to denote the wave propagating in the 
direction of the flow, ($-$) otherwise. Any perturbation can be 
decomposed in the linear approximation onto the acoustic and entropic modes as a 
sum of a pressure perturbation $\delta p =\delta p^+ + \delta p^-$ and an 
entropy perturbation $\delta S$ (\eg Landau \& Lifshitz 1987, Chap. 82).
The linear response of the flow to small perturbations is described by the two 
complex coefficients ${\cal R}_{1},{\cal Q}_{1}$ defined at a position $r_{1}$ 
in the flow as follows
\begin{equation}
{\delta p_1^-\over p_1}={\cal R}_{1}{\delta p_1^+\over p_1}+
{\cal Q}_{1}\delta S_1.\label{rq}
\end{equation}
The coefficient ${\cal R}_{1}$ is directly related to the refraction
of the acoustic flux of an incoming pressure perturbation $\delta p_1^+$ by 
the flow inhomogeneities. Using Eqs.~(\ref{defflux}) 
and (\ref{rq}) with $\delta S=0$, the fraction of refracted acoustic flux is:
\begin{equation}
{F^+_{1}\over F^{-}_{1}} = 
\left({1-\M_{1}\over 1+\M_{1}}\right)^2{\cal R}_{1}^2.
\end{equation}
If there is no incident acoustic wave ($\delta p_1^+=0$), the acoustic
flux $F^{-}_{1}$ propagating against the stream is directly related to the 
amplitude of the incident entropy perturbation, and depend on the 
coefficient ${\cal Q}_{1}$ as follows:
\begin{equation}
F^-_{1}\equiv{\dot M}_0c_{1}^2{(1-\M_{1})^2\over \M_{1}}{\cal Q}_{1}^2
\left({\delta S_{1}\over \gamma }\right)^2.
\end{equation}
Both ${\cal R}_{1}$ and ${\cal Q}_{1}$ depend on the frequency $\omega$ of
the perturbation.
An accurate calculation of ${\cal R}_{1},{\cal Q}_{1}$ in a realistic accretion 
flow is beyond the scope of the present paper, since it would require solving 
the full set of linearized Euler equations. This will be done in the particular 
case of a radial flow in a forthcoming paper (Foglizzo 2000).
The frequency dependence of ${\cal R}_{1}$ can be anticipated by 
remarking that the refraction is negligible for acoustic wave with a 
wavelength much shorter than the scale of the inhomogeneities of the 
flow. In an adiabatic accretion flow, the gravitational potential 
$\Phi\equiv -GM/r$
is responsible for both the acceleration and the heating of the flow. 
The lengthscale of the flow inhomogeneities decreases towards the 
accretor, so that we expect a strong refraction at low 
frequency and a negligible refraction at high frequency.
If the accretion flow is accelerated up to supersonic velocities through a 
sonic surface, acoustic waves with a high enough frequency to 
penetrate in the supersonic region cannot be refracted out: a 
frequency cut-off $\omega_{\rm cut}$ therefore corresponds to the highest 
frequency above which refraction becomes negligible.

\subsection{Estimates of ${\cal R}_{1},{\cal Q}_{1}$ based on the 
compact nozzle approximation\label{Spot}}

We build in Appendix~B a toy model to obtain 
rough estimates of the entropic-acoustic coupling in an accretion flow 
where the mach number increases from $\M_{1}\le 1$ to $1$. The 
refraction point $r_{a}(\omega)$ of acoustic waves of frequency $\omega$
separates the flow into two regions:
\par - a region of propagation $r\ge r_{a}$ where both the 
entropic-acoustic coupling and the acoustic refraction are negligible,
\par - a region $r\le r_{a}$ where entropy perturbations are coupled 
to acoustic waves, and where acoustic waves are refracted.\\
The propagation of acoustic waves in the first region is easily described using 
the conservation of their acoustic flux. The region of coupling is described 
using the "compact nozzle approximation", used for jet nozzles by 
Marble \& Candel (1977). This 
approximation is valid for perturbations with a long enough wavelength in order 
to treat the nozzle as a discontinuity in the flow. The formulae 
obtained by Marble \& Candel (1977) are extended in Appendix~B to 
the case of a flow in a potential $\Phi$. In this academic case, the 
nozzle separates the upstream region characterized by a uniform velocity 
$v_a$, mach number $\M_a<1$ and potential $\Phi_a$, and a downstream 
region characterized by $v_b,\M_b,\Phi_b$. Here again, the wavelength of 
the perturbations is assumed to be longer than the size of the nozzle. 
The compact nozzle approximation enables us to compute the amplitude 
of the acoustic wave $\delta p_a^-/p_a$ propagating upstream 
against the current, triggered by the refraction of an incoming acoustic wave 
$\delta p_a^+/p_a$ or by the advection of an entropy wave $\delta S_a$.\\ 
The following estimates of $|{\cal R}_{1}|,|{\cal Q}_{1}|$ are deduced from
Appendix~B:
\begin{eqnarray}
|{\cal R}_{1}|&\sim& {1+\M_1\over 1-\M_1 }
{\cs^2-v_ac_a\over \cs^2+v_ac_a}\label{r1},\\
|{\cal Q}_{1}|&\sim&
{\M_1^{1\over2}\M_a^{1\over2}\over 1-\M_1}{c_a\over c_1}
{\cs^2-c_a^2\over \cs^2+v_ac_a}.\label{q1}
\end{eqnarray}
Following the argument of Appendix~B, the most efficient entropic-acoustic 
coupling is expected at frequencies close to the refraction cut-off 
$\omega_{\rm cut}$. The maximum value of $|{\cal Q}_{1}|$ is then of the order 
of
\begin{equation}
|{\cal Q}_{1}|\sim 0.3{\cs\over c_{1}}\;\;{\rm if}\;\cs\gg c_{1}.\label{esticomp}
\end{equation}
According to the Bernoulli Eq.~(\ref{Bernoulli}) with a gravitational potential: 
\begin{equation}
\cs^2= 2{\gamma-1\over \gamma+1}\left({GM\over 
\rso}+B\right).\label{cmax}
\end{equation}
The coupling 
coefficient $|{\cal Q}_{1}|$ deduced from Eqs.~(\ref{esticomp}) and (\ref{cmax}) 
can be very large if the sonic radius $\rso$ is close to the accretor, as in 
the radial accretion of a gas with an adiabatic index $\gamma$ close to 5/3 
(Bondi 1952, Foglizzo \& Ruffert 1997). 
 We wish to emphasize the fact that Eq.~(\ref{esticomp}) comes out of a 
rather idealized toy model where both the refraction of acoustic waves and 
their coupling to entropy perturbations occur in a localized "compact" 
region. Nevertheless, we assume it is realistic enough to expect 
$|{\cal Q}_{1}|\gg1$ if $\cs\gg c_{1}$.

\section{Entropy perturbation produced by a sound wave reaching a 
shock\label{sect3}}

The acoustic energy propagating against the flow, released by the processes 
studied above, would escape far from the accretor if no particular boundary 
is met. 
This energy however may feed a cycle if some physical process enables the 
conversion of acoustic energy into entropy
perturbations. A similar cycle occurs in jet nozzles, where sound 
waves reaching the combustion region perturb the combustion and 
produce entropy perturbations. Here we focus on the case of a 
sationary shock satisfying the Rankine-Hugoniot conditions. For the 
sake of simplicity, we restrict ourselves to the case of a shock in 
the plane $y,z$ with incident Mach number $\M_0$ in the $x$ 
direction. The mach number $\M_{\rm sh}<1$ immediately after the 
shock is related to $\M_0>1$ through the classical relation:
\begin{equation}
\M_{\rm sh} ^2=
{2+(\gamma-1)\M_0^2\over2\gamma\M_0^2-\gamma+1}>{\gamma-1\over2\gamma}.
\end{equation}
Let a sound wave $\delta p_{\rm sh}^-$ propagate against the stream along the 
$x$-axis in the subsonic region where the sound speed is $c_{\rm sh}$.
The more general case of any angle of incidence is treated in 
Foglizzo~(2000). The incident sound wave perturbs the shock surface 
and generates both a
reflected sound wave $\delta p_{\rm sh}^+$ and an entropy 
perturbation $\delta S$.
We define the reflexion coefficient $\cpsh$ and the efficiency of 
entropy
production $\cssh$ as follows:
\begin{eqnarray}
\cpsh&\equiv &{\delta p_{\rm sh}^+\over\delta p_{\rm 
sh}^-},\label{defcpsh}\\
\cssh&\equiv & {p_{\rm sh} \delta S\over \delta p_{\rm 
sh}^-}.\label{defcssh}
\end{eqnarray}
A simple calculation in Appendix~C, similar to Landau \& Lifshitz 
(1987, chap.~90), shows that $\cpsh$ and $\cssh$ are real functions of
$\M_{\rm sh}$ and $\gamma$. For a strong shock ($\M_0^2\gg 2/(\gamma-1)$), the 
coefficients $\cpsh,\cssh$ depend only on the adiabatic index $\gamma$:
\begin{eqnarray}
\cpsh&=&-{\gamma^{1\over2}-2^{1\over2}(\gamma-1)^{1\over2}\over
\gamma^{1\over2}+2^{1\over2}(\gamma-1)^{1\over2}},\\
\cssh&=&{2^{3\over2}\over(\gamma-1)^{1\over2}\left\lbrack
\gamma^{1\over2}+2^{1\over2}(\gamma-1)^{1\over2}\right\rbrack}.
\end{eqnarray}

\section{Global instability\label{sect4}}

 At a given frequency $\omega$, let us define $\tau_+,\tau_E$ as respectively 
the time of acoustic propagation and the advection time from the shock to the 
refraction point, and $\tau_-$ the time of 
acoustic propagation from the refraction poin to the shock. 
$\tau_{AA}\equiv\tau_-+\tau_+$ is the duration of the purely 
acoustic cycle, while $\tau_{EA}\equiv\tau_-+\tau_E$ is the duration 
of the entropic-acoustic cycle. 

\subsection{Instability of the entropic-acoustic cycle\label{globmec}}

\begin{figure}
\psfig{file=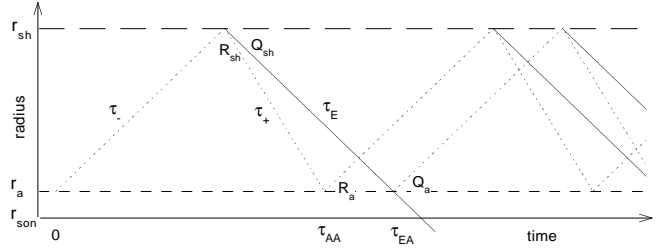,width=\columnwidth}
\caption[]{Schematic trajectories of acoustic (dotted lines) and 
entropic (full lines) perturbations forming the entropic-acoustic and 
the purely acoustic parts of the cycle. Acoutic waves 
are reflected against the shock and refracted by the flow 
gradients with the respective efficiencies $\cpsh$ and ${\cal R}_{a}$.
$r_{a}>\rso$ is the refraction radius. Entropy perturbations are created by 
the interaction of acoustic waves with the shock
with the efficiency $\cssh$. Their advection triggers outgoing acoustic 
waves with the efficiency ${\cal Q}_{a}$. Acoustic propagation times 
$\tau_{-},\tau_{+}$ and entropic advection time $\tau_{E}$  
are also illustrated.
}
\label{figphi0}
\end{figure}
A first estimate of the entropic-acoustic instability is obtained by 
neglecting the effect of the purely acoustic cycle.
The adiabatic advection of an entropy perturbation of amplitude 
$\delta S$ into regions of higher enthalpy produces acoustic waves 
$\delta p_{\rm sh}^-/p_{\rm sh}$ estimated in Sect.~\ref{Sconvent}.
When these acoustic waves reach the shock, new entropy perturbations 
are created and were estimated in Sect.~\ref{sect3}. This leads us to 
define the global efficiency ${\cal Q}$ of the entropic-acoustic cycle as 
follows: 
\begin{equation}
{\cal Q}\equiv {\cal Q}_{1}\cssh,
\end{equation}
As for classical problems of modes in \eg acoustic cavities, the integral phase 
condition for the existence of an entropic-acoustic mode constrains the real 
part $\omega_{r,n}$ of its frequency:
\begin{equation}
\omega_{r,n} ={n\pi\over\tau_{EA}},\label{reomega}
\end{equation}
where the integer $n$ must be even if ${\cal Q}>0$ (\eg in an  accretion 
flow), and odd otherwise (\eg in a jet nozzle).
Note that the duration $\tau_{EA}$ of the entropic-acoustic cycle also
depends on the frequency through the position of the refraction point. 
The condition of instability of the entropic-acoustic cycle is $|{\cal 
Q}|>1$. This leads us to define the global growth (or damping) rate 
$\omega_{i,n}$ associated to the mode $n$:
\begin{equation}
\omega_{i,n} = {1\over\tau_{EA}}\log |{\cal Q}|.\label{imomega}
\end{equation}

\subsection{Which accretion flows are unstable ?} 

On the one hand, we know from Sect.~\ref{Spot} that the coefficient 
$|{\cal Q}_{1}|\gg1$ if $\cs\gg c_{\rm sh}$. On the other hand, we show in 
Appendix~D that the coefficient $\cssh$ may damp the instability if the shock is 
too weak. Thus the global efficiency $|{\cal Q}|$ exceeds unity for flows 
such that $\cs\gg c_{\rm sh}$ with a strong shock, as in the Bondi-Hoyle 
accretion with $\gamma$ close to $5/3$ at high Mach number.
Following the compact approximation, the frequency of the most unstable mode 
is close to the refraction cut-off $\omega_{\rm cut}$, which corresponds 
to a very high order $n_{\rm max}$ if $\cs\gg c_{\rm sh}$:
\begin{equation}
n_{\rm max}\sim {\omega_{\rm cut}\tau_{EA}\over \pi}\gg1.
\end{equation}
Eq.~(\ref{imomega}) indicates that the growth rate depends 
logarithmically on $|{\cal Q}|$, so that the maximum growth 
rate is always of the order of the inverse of the duration $\tau_{EA}$ of a 
whole entropic-acoustic cycle if $|{\cal Q}|\gg 1$. Using 
Eq.~(\ref{esticomp}), 
\begin{equation}
\omega_{i,\rm max} \sim {1\over\tau_{EA}}\log {\cs\over c_{\rm sh}}.
\label{imomegamax}
\end{equation}

\subsection{Contribution of the purely acoustic cycle}

\begin{figure}
\psfig{file=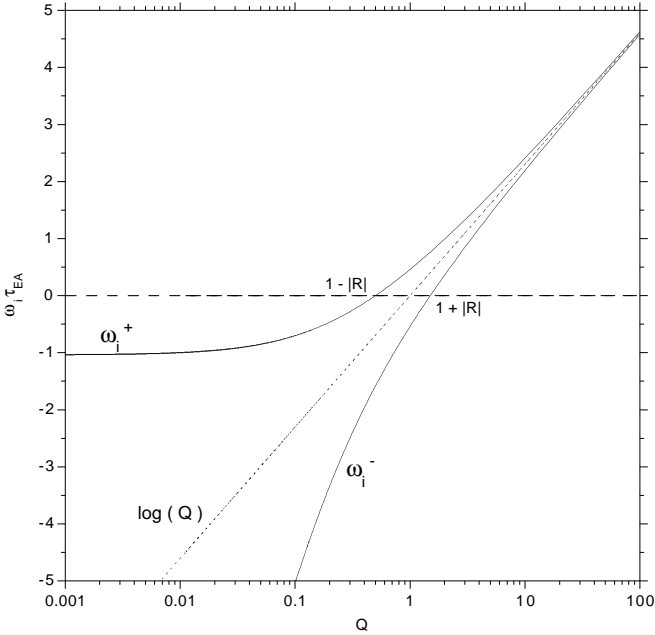,width=\columnwidth}
\caption[]{Growth rate as a function of the global efficiency ${\cal Q}$, 
for $\tau_{AA}/\tau_{EA}=2/3$, $|{\cal R}|=0.5$. The dotted line corresponds to 
the entropic-acoustic cycle alone ${\cal R}=0$. The full curves give the upper
and lower bounds $\omega_i^\pm$ of the growth rate when the contribution of the 
purely acoustic cycle is taken into account. $\omega_i^\pm=0$ for 
$|{\cal Q}|=1\mp|{\cal R}|$. Eigenmodes are damped for $\omega_i<0$, and 
unstable for $\omega_i>0$.
}
\label{figminmax}
\end{figure}
A refined estimate of the growth rate and frequency of the 
instability should take into account the sound waves $\delta p_{\rm 
sh}^+$ produced at the shock with the efficiency $\cpsh$, and their 
refraction near the sonic radius with the efficiency ${\cal R}_{1}$. 
The global efficiency ${\cal R}$ of the purely acoustic cycle is:
\begin{equation}
{\cal R}\equiv {\cal R}_{1}\cpsh.
\end{equation}
The eigenfrequencies must satisfy the following equation, deduced from 
Eqs.~(\ref{rq1}) and (\ref{defcpsh}-\ref{defcssh}):
\begin{equation}
{\cal Q}\e^{i\omega\tau_{EA}}+
{\cal R}\e^{i\omega\tau_{AA}}=1.\label{dispersion}
\end{equation}
The phase relation between the purely acoustic and the 
entropic-acoustic cycles
is rather complicated (see Eq.~\ref{phase}). 
The calculation of the eigenfrequencies 
requires to know the exact dependence of ${\cal Q},{\cal R}, 
\tau_{EA},\tau_{AA}$ on the frequency. We show in Appendix~E
that the growth rate $\omega_{i}$ is bounded to the interval 
$[\omega_i^-,\omega_i^+]$ satisfying the equation:
\begin{equation}
|{\cal Q}|\e^{-\omega_i^\pm\tau_{EA}}
\pm|{\cal R}|\e^{-\omega_i^\pm\tau_{AA}}=1.\label{wminmax}
\end{equation}
The solutions $\omega_i^\pm$ of this equation are plotted as 
functions of $\cal Q$ in Fig.~\ref{figminmax}. 
All the modes are unstable for 
$|{\cal Q}|>1+|{\cal R}|$, while all of them are stable for $|{\cal 
Q}|<1-|{\cal R}|$. In the range $1-|{\cal R}|<|{\cal Q}|<1+|{\cal R}|$, 
both stable and unstable modes may be found, depending on the phase of 
the acoustic cycle. A first order 
calculation of the contribution of the acoustic cycle when $|{\cal 
R}|/|{\cal Q}|^{\tau_{AA}/\tau_{EA}}\ll1$ leads to:
\begin{eqnarray}
\omega_{r,n}&\sim& {1\over\tau_{EA}}\left\lbrack n\pi - 
{{\cal R}\over|{\cal Q}|^{\tau_{AA}\over\tau_{EA}}}
\sin\left(n\pi{\tau_{AA}\over\tau_{EA}}\right)\right\rbrack,
\label{omegarn}\\
\omega_{i,n}&\sim& {1\over\tau_{EA}}\left\lbrack\log |{\cal Q}|  +
{{\cal R}\over|{\cal Q}|^{\tau_{AA}\over\tau_{EA}}}
\cos\left(n\pi{\tau_{AA}\over\tau_{EA}}\right)\right\rbrack,
\label{omegain}
\end{eqnarray} 
where $n$ is even if ${\cal Q}>0$, and odd otherwise.
The acoustic cycle can be stabilizing or destabilizing depending on 
the phase $n\pi\tau_{AA}/\tau_{EA}$. This phase dependence results in a 
wiggling in the spectrum of the eigenmodes, bounded by $\omega_i^\pm$.
  
\subsection{Comparison with the particular case of jet nozzles}

The velocity and enthalpy profiles are quite different in a jet 
nozzle and in an accretion flow. The gas is accelerated in a nozzle 
at the expense of the enthalpy, whereas gas flowing into a 
gravitational well $\Phi\equiv-GM/r$ is both accelerated and heated. 
The limited variation of enthalpy in a nozzle, deduced from the 
Bernoulli equation with $\Phi\equiv 0$,
\begin{equation}
{\cs^2\over c_{\rm sh}^{2}}={2+(\gamma-1)\M_{\rm sh}^2\over\gamma+1}<1,
\end{equation}
results in a moderate efficiency of the extraction of acoustic energy 
from the advection of entropy perturbations ($|{\cal Q}|<1$).
Nevertheless, the combustion zone plays a much more active role than 
the shock described above, since the energy input due to the 
chemistry of the combustion enables the actual efficiency $|{\cal Q}|$ 
to exceed unity. In this respect, the "rumble" instability of ramjets 
(Abouseif, Keklak \& Toong 1984) is different from the entropic-acoustic 
instability described here. Both are based, however, on the cycle of 
entropic and acoustic perturbations in the subsonic region of the 
converging flow. 
In an accretion flow, the efficiency $|{\cal Q}|$ increases with the 
frequency up to the refraction cut-off, thus favouring a high 
frequency instability ($n_{\rm max}\gg1$). By contrast, the most 
unstable modes in Abouseif, Keklak \& Toong (1984) are the 
low frequency modes. 

\section{Discussion\label{Sconclusion}}

We have described how the interplay of entropic and acoustic waves in 
a subsonic flow between a shock and a sonic point
might lead to an entropic-acoustic linear instability. The 
ingredients for this type of instability are 
\par(1) an accelerated subsonic flow with a significant increase of 
sound speed (\ie of temperature),
\par(2) an outer boundary, such as a shock, capable of converting 
acoustic waves into entropy perturbations.\\
The stability criterion is directly related to the calculation of a 
global parameter ${\cal Q}$ describing the coupling of 
entropy and acoustic waves in the flow. The effect of the purely acoustic 
cycle is negligible if $|{\cal Q}|\gg1$, but becomes important near 
marginal stability (Eq.~\ref{wminmax}). Although a detailed stability 
calculation is required for any particular flow topology, general arguments 
based on the compact nozzle approximation suggest that the entropic-acoustic 
cycle is potentially unstable if the sound speed at the sonic point 
significantly exceeds the sound speed immediately after the shock 
(Eq.~\ref{imomegamax}). The two timescales involved in the entropic-acoustic 
instability are:
\par- the period of the most unstable pertubations, of the order of 
the refraction cut-off,
\par- the growth time, comparable to the time needed to advect 
an entropy perturbation from the shock to the sonic radius.\\

\subsection{Astrophysical applications}

The entropic-acoustic instability might be at work in several astrophysical 
situations: 
as already mentionned in the introduction, the supersonic motion of a compact 
star in a gas leads to the formation of a bow shock ahead of the 
accretor. The development of the entropic-acoustic instability in the subsonic 
region between the shock and the sonic surface is very likely, 
since $\cs/c_{\rm sh}\gg1$ in these flows (\eg Ruffert 1996 and references 
therein). A detailed analysis will be presented in a forthcoming paper.\\
The acoustic-entropic instability might also be present in some 
accretion discs. Most stability calculations concerning ADAFs focussed 
on the {\it local} thermal stability of the disc (Narayan \& Yi 1995, 
Abramowicz \etal 1995, Kato \etal 1996, 1997, Wu \& Li 1996). By 
contrast, the entropic-acoustic instability is a {\it global} 
instability which depends on the outer boundary condition of the flow.
The excitation of acoustic waves from an advected 
entropy perturbation in an ADAF has already been observed in 
numerical simulations by Manmoto \etal (1996), which include the 
additional effects of radiative cooling (bremmstrahlung) and viscous 
heating. Their work was concerned with the time delay problem of 
Cygnus X-1 and did not involve an outer boundary condition capable of 
closing the entropic-acoustic cycle. Some global disc solutions match a 
thin disc to an inner ADAF through a shock (Lu, Gu \& Yuan 1999). The 
entropic-acoustic cycle could then occur between this shock and the 
inner sonic surface. On a larger scale, the gas falling from the 
Lagrangian point $L_{1}$ onto the external radius of an accretion disc might 
also produce a shock capable of closing the entropic-acoustic cycle.
These possible applications deserve a careful analysis which is 
beyond the scope of the present paper.

\subsection{Towards a 3-D description of the instability}

Our one-dimensional description assumed that all the outgoing 
acoustic waves reach the shock, and that all the induced 
perturbations of entropy reach the sonic surface. Depending on the 
topology of the flow, possible losses from the entropic-acoustic 
cycle must be taken into account. In the case of Bondi-Hoyle 
accretion, outgoing acoustic waves reaching the shock out of the 
accretion cylinder produce entropy perturbations which will not be 
accreted, thus diminishing the effective efficiency ${\cal Q}$ of the 
cycle. A 3-D description of the entropic-acoustic cycle should also 
take into account the effect of vorticity perturbations,
which are naturally created when an acoustic wave perturbs a shock 
with a non vanishing angle of incidence. Like the entropy 
perturbations, the advection of a vorticity perturbation in a 
converging flow excites acoustic waves
(\eg Howe 1975). Their destabilizing effect will be studied in the 
particular case of radial accretion (Foglizzo 2000).

\subsection{Observable consequences}

Understanding the fundamental mechanism of linear instability is only the 
very first step towards a possible identification of this mechanism in the 
astrophysical observations. 
According to our linear description of the entropic-acoustic 
instability, we expect to observe significant variations of the mass 
accretion rate and displacements of the shock surface as the acoustic 
energy increases between the shock and the sonic surface. The 
determination of the spectrum of eigenfrequencies should give 
valuable informations about the timescale of the most unstable modes.
Numerical simulations should be the most direct way to describe the ultimate 
non-linear development of this instability. We saw that a 
correct description of the instability requires an accurate 
propagation of acoustic and entropy waves between the shock and the 
sonic radius. This could be a serious numerical difficulty for flows 
where $\cs\gg c_{\rm sh}$, since the frequency of the most unstable 
perturbations could exceed the growth rate by several orders of 
magnitude.  The numerical simulations of Bondi-Hoyle-Lyttleton accretion 
(\eg Ruffert 1995) suggest that this instability might lead to some kind of 
turbulence. This could also be particularly promising if the same mechanism 
also applies to accretion discs. 

\acknowledgements The authors are grateful to Dr. M. Ruffert for 
useful discussions, and to an anonymous referee for his constructive 
comments.

\appendix

\section{Energy release of a localized perturbation of entropy 
through a nozzle \label{Aenerg}}

We consider in the region I an entropy perturbation $\delta S$, 
corresponding to a perturbation of mass $\delta m$ initially in 
pressure 
equilibrium with the surrounding gas ($\delta p_1=0$, $\delta v_1=0$). 
The time scale $\tau$ associated to this perturbation of entropy is 
related to 
its length $l$:
\begin{eqnarray}
\tau&=&{l\over v},\\
\delta m &=& {\delta\rho\over\rho}{\dot M}_0\tau ,
\end{eqnarray}
where ${\dot M}_0$ is the unperturbed mass flux.
The entropy perturbation is conserved when advected:
\begin{equation}
\delta S\equiv{1\over\gamma-1}\log \left(1+{\delta p\over p} 
\right)\left(1+{\delta\rho\over\rho}\right)^{-\gamma},\label{entropy}
\end{equation}
The advected perturbations of density and sound speed 
($c^2\equiv \gamma p /\rho$) follow from Eqs.~(\ref{entropy}):
\begin{eqnarray}
{\delta\rho_1\over\rho_1}&=&
\exp\left(-{\gamma-1\over\gamma}\delta S\right) -1 ,\label{drho}\\
{\delta c^2_1\over c^2_1}&=&
\exp\left({\gamma-1\over\gamma}\delta S\right) -1.\label{dc}
\end{eqnarray}
The energy flux $F_0+\delta F_1$ in the perturbation of entropy is 
defined according to Landau \& Lifshitz (1987, Chap. 6), and deduced 
from Eqs.~(\ref{Bernoulli}), (\ref{drho}) and (\ref{dc}):
\begin{eqnarray}
\delta  F_1 &\equiv& ({\dot M}_0+\delta {\dot M})(B_0+\delta B) - 
{\dot M}_0B_0,\\
&=&{\dot M}_0\left({v_1^2\over2}+\Phi_1\right)\left\lbrack
\exp\left(-{\gamma-1\over\gamma}\delta S\right) 
-1\right\rbrack.\label{de1}
\end{eqnarray} 
Replacing the index $1$ by the index $2$ gives the energy flux 
$F_0+\delta F_2$ in the entropy perturbation once it has reached 
pressure 
equilibrium with the region II. The total energy $\delta E_{12}$ 
released 
during the advection through the nozzle is therefore:
\begin{eqnarray}
\delta E_{12}&\equiv &(\delta F_1-\delta F_2)\tau,\\
&=& {\dot M}_0\tau(h_2-h_1)\left\lbrack
\exp\left(-{\gamma-1\over\gamma}\delta S\right) -1
\right\rbrack,\label{deltae2}\\
&=&(h_2-h_1)\delta m.
\end{eqnarray}

\section{The compact nozzle in a potential $\Phi(r)$\label{Acompact}}
\subsection{Conservation equations} 
The variations $\delta {\dot M}^\pm$ and ${\delta B}^\pm$ of the mass 
flux ${\dot M}$ and Bernoulli parameter $B$ within a pressure 
perturbation $\delta p^\pm$ ($\pm1$ depending on the direction of 
propagation of the perturbation) is to first order:
\begin{eqnarray}
{\delta {\dot M}^\pm\over{\dot M}}&=&{\delta p^\pm\over\gamma 
p}{\M\pm1\over\M},\label{dmp}\\
{\delta B}^\pm&=&c^2(1\pm\M){\delta p^\pm\over\gamma p}.\label{dpb}
\end{eqnarray}
In an entropy perturbation $\delta S$ in pressure equilibrium with 
its surrounding, the corresponding variations $\delta {\dot M}^e$ and 
${\delta B}^e$ are as follows:
\begin{eqnarray}
{\delta {\dot M}^e\over{\dot M}}&=&-{\gamma-1\over\gamma}\delta 
S,\label{dms}\\
{\delta B}^e&=&{c^2\over\gamma}\delta S.\label{dbs}
\end{eqnarray}
Writing the conservation of the perturbed entropy, mass flux and 
Bernoulli parameter across a subsonic nozzle leads to the equations 
for the subcritical nozzle:
\begin{eqnarray}
\delta S_a&=&\delta S_b,\\
\delta {\dot M}_a^+ +\delta {\dot M}_a^- + \delta {\dot M}_a^e &=&
\delta {\dot M}_b^+ +\delta {\dot M}_b^- + \delta {\dot M}_b^e,\\
\delta B_a^+ +\delta B_a^- + \delta B_a^e &=&
\delta B_b^+ +\delta B_b^- + \delta B_b^e.
\end{eqnarray}
The case of a supercritical nozzle requires one more equation 
describing the effect of perturbations on the sonic surface. The 
Bernoulli equation can be written at the sonic point $\rso$ in a flow 
tube of entropy $S$ with mass flux ${\dot M}$ and local cross section 
${\cal A}_s$:
\begin{equation}
B-\Phi(\rso)={\gamma\over2}{\gamma+1\over\gamma-1}\left\lbrack
{{\dot M}\e^S\over \gamma^{1\over2}{\cal A}_s} 
\right\rbrack^{2(\gamma-1)\over\gamma+1}.\label{bersonic}
\end{equation}
Differentiating the Bernoulli equation and the mass conservation at 
the sonic point leads to
\begin{equation}
\cs^2{\delta {\cal A}\over {\cal A}_s}=\delta \Phi(\rso).
\end{equation}
Combining this equation with the differential of Eq.~(\ref{bersonic}) 
provides us with the additional equation required to solve the case 
of a supercritical nozzle:
\begin{equation}
{\delta {\dot M}\over {\dot M}}+\delta S-{\delta B\over \cs^2}=0.
\end{equation}

\subsection{Incident acoustic wave propagating with the stream}
Subcritical nozzle:
\begin{eqnarray}
{\delta p_a^-\over\gamma p_a}&=&{1+\M_a\over1-\M_a}
{v_bc_b-v_ac_a\over v_bc_b+v_ac_a}{\delta p_a^+\over\gamma 
p_a},\label{subpa}\\
{\delta p_b^+\over\gamma p_b}&=&{1+\M_a\over1+\M_b}
{2\M_b c_a^2\over v_bc_b+v_ac_a}{\delta p_a^+\over\gamma 
p_a}.\label{subp2}
\end{eqnarray}
Supercritical nozzle:
\begin{eqnarray}
{\delta p_a^-\over\gamma p_a}&=&{1+\M_a\over1-\M_a}
{\cs^2-v_ac_a\over \cs^2+v_ac_a}{\delta p_a^+\over\gamma p_a},
\label{suppa}\\
{\delta p_b^+\over\gamma p_b}&=&{1+\M_a\over1+\M_b}{c_a^2\over c_b^2}
{\cs^2+v_bc_b\over \cs^2+v_ac_a}{\delta p_a^+\over\gamma p_a},\\
{\delta p_b^-\over\gamma p_b}&=&{1+\M_a\over1-\M_b}{c_a^2\over c_b^2}
{\cs^2-v_bc_b\over \cs^2+v_ac_a}{\delta p_a^+\over\gamma p_a}.
\end{eqnarray}

\subsection{Incident entropy perturbation advected with the stream}
Subcritical nozzle:
\begin{eqnarray}
{\delta p_a^-\over\gamma p_a}&=&{\M_a\over1-\M_a}
{c_b^2-c_a^2\over v_bc_b+v_ac_a}{\delta S\over\gamma},\label{dpa-}\\
{\delta p_b^+\over\gamma p_b}&=&-{\M_b\over1+\M_b}
{c_b^2-c_a^2\over v_bc_b+v_ac_a}{\delta S\over\gamma}.
\end{eqnarray}
Supercritical nozzle:
\begin{eqnarray}
{\delta p_a^-\over\gamma p_a}&=&{\M_a\over1-\M_a}
{\cs^2-c_a^2\over \cs^2+v_ac_a}{\delta S\over\gamma},\label{dpas}\\
{\delta p_b^+\over\gamma p_b}&=&-{\cs^2-c_a^2\over2(\cs^2+v_ac_a)}
\left\lbrack 1 +
{c_a^2\over c_b^2}{1+\M_a\over1+\M_b}{c_b^2-\cs^2\over\cs^2-c_a^2}
\right\rbrack{\delta S\over\gamma},\nonumber\\
& &\\
{\delta p_b^-\over\gamma p_b}&=&-{\cs^2-c_a^2\over2(\cs^2+v_ac_a)}
\left\lbrack 1 +
{c_a^2\over c_b^2}{1+\M_a\over1-\M_b}{c_b^2-\cs^2\over\cs^2-c_a^2}
\right\rbrack{\delta S\over\gamma}.\nonumber\\
& &
\end{eqnarray}

\subsection{Acoustic flux $F_a^-$ propagating upstream \label{Scompact}}

Since our calculation is based on a linear approximation, we may define 
efficiencies ${\cal R}_{a},{\cal Q}_{a}$ of the compact nozzle as follows
\begin{equation}
{\delta p_a^-\over p_a}={\cal R}_{a}{\delta p_a^+\over p_a}+
{\cal Q}_{a}\delta S_a,\label{rqa}
\end{equation}
where ${\cal R}_{a},{\cal Q}_{a}$ are easily deduced from 
Eqs.~(\ref{subpa}) and (\ref{dpa-}) or Eqs.~(\ref{suppa}) and (\ref{dpas}),
depending on whether the nozzle is subcritical ($\M_b<1$), or supercritical 
($\M_b>1$). The pressure perturbation $\delta p_a^-$ generated upstream in the 
case of a supercritical nozzle (Eqs.~\ref{suppa}-\ref{dpas}) 
corresponds to the limit of Eqs.~(\ref{subpa}-\ref{dpa-}) when 
$c_b=\cs$, the sound speed at the sonic point. This is consistent with 
the fact that no information about 
the fluid properties beyond the sonic point may reach the subsonic 
region upstream. Note from Eq.~(\ref{dpas}) that the sign of ${\cal Q}_{a}$ 
depends on whether the sound speed of the accelerated gas increases or 
decreases. The Bernoulli equation implies that ${\cal Q}_{a}<0$ in a 
classical nozzle ($\Phi\equiv0$), while ${\cal Q}_{a}>0$ in an accretion flow.
The flux $F_a^-$ of acoustic energy propagating 
against the stream when an acoustic wave carrying the acoustic flux $F_a^+$ 
(with $\delta S=0$) propagates into a supercritial nozzle is deduced from 
Eqs.~(\ref{defflux}) and (\ref{suppa}):
\begin{equation}
{\bar{\cal R}_{a}}^2\equiv {F_a^-\over F_{a}^+} =
\left({\cs^2-v_ac_a\over \cs^2+v_ac_a}\right)^2,
\label{fluxpress}
\end{equation}
where the refraction efficiency ${\bar{\cal R}_{a}}^2$ is defined in 
terms of the acoustic energy, while ${\cal R}_{a}^2$ was defined in 
Eq.~(\ref{rqa}) in terms of the amplitude of pressure perturbations.
It is remarkable that the refraction $F_a^-/F_a^+$ is 
independent of the frequency of the incoming acoustic wave. This is true 
only at low frequency, as long as the compact approximation is satisfied. 
Refraction is expected to become negligible ($F_a^-/F_a^+\ll1$) at high 
frequency, when the wavelength of an incoming acoustic wave is much shorter 
than the lengthscale $\lambda$ of the flow inhomogeneities (\ie the nozzle 
size). The refraction cut-off $\omega_{\rm cut}$ is defined as the maximum 
frequency leading to a significant refraction (\eg $F_a^-/F_a^+=1/2$).\\
The flux $F_a^-$ of acoustic energy propagating against the stream when an 
entropy perturbation $\delta S$ (with $\delta p^+=0$) is advected into a 
supercritical nozzle is deduced from Eqs.~(\ref{defflux}) and (\ref{dpas}):
\begin{equation}
F_a^-={\dot M}_0 v_ac_a\left({\cs^2-c_a^2\over \cs^2+v_ac_a}\right)^2
\left({\delta S\over\gamma}\right)^2.\label{flux3}
\end{equation}
Here again, the acoustic flux $F_a^-$ is 
independent of the frequency of the incoming entropy wave, as long as the 
compact approximation is satisfied. 
Numerical calculations by Marble \& Candle (1977) show that the 
efficiency of sound emission from an entropy perturbation also 
decreases to zero if its wavelength is much shorter than $\lambda$. 
Eq.~(\ref{flux3}) is rewritten using the refraction coefficient $\rfr$:
\begin{eqnarray}
F_a^-&=& {\dot M}_0\cs^2
 \rfr^2\left({1-\rfr\over1+\rfr}\right)
\left({\cs^2-c_a^2\over\cs^2-c_av_a}\right)^2
\left({\delta S\over\gamma}\right)^2.
\end{eqnarray}
The ratio $(\cs^2-c_a^2)/(\cs^2-c_av_a)$ is bounded by
\begin{eqnarray}
{\cs^2-c_a^2\over\cs^2-c_av_a}&\sim&1 \;\;{\rm for}\;\;
c_a\ll\cs,\\
&\sim& \left(1 + {{\dot\M}_{\rm 
son}\cs\over2{\dot\cs}}\right)^{-1}\;\;{\rm for}\;\;
c_a\sim \cs,
\end{eqnarray}
where ${\dot\M}_{\rm son},{\dot\cs}$ are the derivatives of the mach 
number and sound speed at the sonic point. Schematically, $\rfr$ is 
constant for $\omega\ll\omega_{\rm cut}$ and decreases to zero
for $\omega\gg\omega_{\rm cut}$. The maximal value of 
$\rfr^2(1-\rfr)/(1+\rfr)\sim 0.09$ is reached for $\rfr\sim 0.6$, \ie 
$v_{a}c_{a}\sim \cs^2/4$. We conclude that the acoustic flux propagating 
against the stream when $F_a^+=0$ is bounded by:
\begin{equation}
F_a^-\le 0.09\;{\dot M}_0\cs^2
\left({\delta S\over\gamma}\right)^2.
\end{equation}

\subsection{Combination of a slowly accelerated flow and a compact 
nozzle\label{Scombi}}

Let us build a toy model of a flow $\M_{1}\le\M\le 1$, especially 
designed to allow a simple calculation of the interplay of acoustic and 
entropic perturbations of frequency $\omega$. This flow is composed of:
\par - a region of propagation $\M_{1}\le\M\le \M_{a}$ where both the 
entropic-acoustic coupling and the acoustic refraction at the 
frequency $\omega$ are negligible, and where the flow is slowly 
accelerated,
\par - a compact region $\M_{a}\le\M\le 1$ of coupling, approximated as 
a supercritical compact nozzle.\\
The index $a$ notes the entrance of the compact nozzle.
These two approximations are compatible if the lengthscale of the 
flow inhomogeneities is small compared to the wavelength of the 
perturbations in the compact region, and large 
compared to this wavelength in the propagation region.
The real coefficients ${\cal P}_\pm,{\cal P}_E$ describe the amplitude variation
of the pressure and entropy perturbations during their propagation 
between $\M_1$ and $\M_a$:
\begin{eqnarray}
{\delta p_a^+\over p_a}&\equiv&{\cal P}_+{\delta p_1^+\over p_1}
e^{i\omega\tau_+},\\
{\delta p_1^-\over p_1}&\equiv&{\cal P}_-{\delta p_a^+\over p_a}
e^{i\omega\tau_-},\\
\delta S_a&\equiv&{\cal P}_E\delta S_1e^{i\omega\tau_E}.
\end{eqnarray}
The values of ${\cal P}_\pm$ are directly deduced from the 
conservation of the acoustic flux from $\M_{a}$ to $\M_{1}$ 
($F_1^\pm\sim F_a^\pm$), and ${\cal P}_E$ is deduced from the conservation 
of entropy:
\begin{eqnarray}
{\cal P}_+&=&{1+\M_1\over 1+\M_a}{\M_a^{1\over2}c_1\over\M_1^{1\over2}c_a}
,\label{proplus}\\
{\cal P}_-&=&{1-\M_a\over 1-\M_1}{\M_1^{1\over2}c_a\over\M_a^{1\over2}c_1}
,\\
{\cal P}_E&=&1.\label{propent}
\end{eqnarray}
Combining Eqs.~(\ref{rqa}) and (\ref{proplus}-\ref{propent}), we 
obtain the amplitude $\delta p_1^-$ of the acoustic waves propagating 
against the stream at the position $\M_1$, as a function of the incident 
perturbations $\delta p_1^+,\delta S_1$:
\begin{eqnarray}
{\cal R}_1&=&{\cal P_-}{\cal P_+}{\cal R}_{a}e^{i\omega\tau_{AA}}\\
{\cal Q}_1&=&{\cal P_-}{\cal P}_{E}{\cal Q}_{a}
e^{i\omega\tau_{EA}}.\label{rq1}
\end{eqnarray}
Eqs.~(\ref{r1}-\ref{q1}) are deduced from Eqs.~(\ref{suppa}), (\ref{dpas}) and 
Eqs.~(\ref{proplus}) to (\ref{propent}). From the discussion of 
Sect.~\ref{Scompact}, the strongest acoustic flux $F_{1}^-$ is 
obtained for $\rfr\sim 0.6$

\section{Reflexion of a sound wave at a shock\label{appsh}}

Consider an adiabatic shock with incident Mach number $\M_0$ in the 
$x$ direction. Let a sound wave propagate against the stream in the subsonic
region where the sound speed is $c_{\rm sh}$, with wavevector 
$k_{-}$ in the frame moving with the fluid. 
Its frequency $\omega$ in the frame stationary with respect to the 
shock obeys the following dispersion relation 
(\eg Landau \& Lifschitz, 1986, chap. 68):
\begin{equation}
(\omega-k v_{\rm sh})^2 = c_{\rm sh}^2k^2.\label{dispson}
\end{equation}
The two roots $k_{\mp}$ of this equation correspond to the 
incident and reflected acoustic waves.
The incident acoustic wave perturbs the shock surface and generates both 
a reflected acoustic wave and an entropy perturbation.
The pressure perturbation $\delta p_{\rm sh}$ and the entropy perturbation 
$\delta S$ are obtained by writing the jump conditions for an adiabatic 
shock perturbed with the velocity $\Delta v$ in the $x$ direction:
\begin{eqnarray}
{\delta p_{\rm sh}\over p_{\rm sh}} &=&-{4\gamma\M_0^2\over 
2\gamma\M_0^2-\gamma+1}
{\Delta v\over v_0},\\
{\delta v_{\rm sh}\over v_{\rm 
sh}}&=&{2(1+\M_0^2)\over2+(\gamma-1)\M_0^2}
{\Delta v\over v_0},\\
{\delta S}&=&-{4\gamma(\M_0^2-1)^2\over
(2+(\gamma-1)\M_0^2)(2\gamma\M_0^2-\gamma+1)}{\Delta v\over v_0}.
\end{eqnarray}
The perturbation is decomposed onto the ingoing and outgoing acoustic 
waves, and the advected entropy perturbation
\begin{eqnarray}
\delta p_{\rm sh}&=&\delta p_{\rm sh}^- + \delta p_{\rm sh}^+,\\
\delta v_{\rm sh}&=&\delta  v_{\rm sh}^- +\delta v_{\rm sh}^+ 
+\delta v_{\rm sh}^e,
\end{eqnarray}
In addition to the dispersion equation (\ref{dispson}) for acoustic 
waves, the acoustic and entropy perturbations are described by:
\begin{eqnarray}
(\omega-k_{\pm})\delta v_{\rm sh}^\pm &=& k_{\pm}{c_{\rm 
sh}^2\over\gamma}
{\delta p_{\rm sh}^\pm\over p_{\rm sh}},\\
\omega &=& k_{e}v_{\rm sh},\\
k_{e}\delta v^e_{\rm sh}&=&0.\label{divv}
\end{eqnarray}
After some algebra with Eqs.~(\ref{dispson}) to (\ref{divv}), we 
obtain:
\begin{eqnarray}
\cpsh&=& -\left({1-\M_{\rm sh}\over1+\M_{\rm sh}}\right)^2
{3-\gamma-2(\gamma-1)\M_{\rm sh}\over3-\gamma+2(\gamma-1)\M_{\rm 
sh}}<0
,\label{cpsh}\\
\cssh&=&{(1-\M_{\rm sh})^2\over \M_{\rm sh}}{4\over
3-\gamma+2(\gamma-1)\M_{\rm sh}}>0.
\label{cssh}
\end{eqnarray}

\begin{eqnarray}
\rfsh&\equiv&
-\left\lbrack{1-\M_{\rm sh}\over1+\M_{\rm sh}}\right\rbrack
\left\lbrack{3-\gamma-2(\gamma-1)\M_{\rm sh}\over
3-\gamma+2(\gamma-1)\M_{\rm sh}}\right\rbrack,
\end{eqnarray}
where $\rfsh^2$ is the reflection coefficient of acoustic waves defined 
in terms of energy flux rather than amplitudes.
For a strong shock, $\rfsh$ is an increasing function of $\gamma$ 
($\rfsh\sim-0.14$ for $\gamma=5/3$):
\begin{equation}
\rfsh=-{2^{1\over2}-\gamma^{1\over2}(\gamma-1)^{1\over2}
\over2^{1\over2}-\gamma^{1\over2}(\gamma-1)^{1\over2}}.
\end{equation}

\section{Estimates of the global efficiencies ${\cal Q}$ and ${\cal R}$ 
based on the compact nozzle approximation}

\begin{eqnarray}
{\cal R}&\equiv &{\cal P}_-{\cal P}_+{\cal R}_{a}\cpsh,\\
&=&\rfr \rfsh.
\end{eqnarray}
${\cal R}$ depends on the frequency through $\rfr$: ${\cal R}\sim 
\rfsh$ for $\omega\ll\omega_{\rm cut}$, and 
${\cal R}\sim 0$ for $\omega\gg\omega_{\rm cut}$.
\begin{eqnarray}
{\cal Q}&\equiv &{\cal P}_-{\cal P}_E{\cal Q}_{a}\cssh,\\
&=&\left({v_ac_a\over v_{\rm sh}c_{\rm sh}}\right)^{1\over2}
{\cs^2-c_a^2\over\cs^2+v_ac_a}\qfq,\\
\qfq&\equiv&
{4\over\M_{\rm sh}^{1\over2}}
\left\lbrack{1-\M_{\rm sh}\over3-\gamma+2(\gamma-1)\M_{\rm sh}}
\right\rbrack.
\end{eqnarray}
The efficiency ${\cal Q}$ decreases to zero as $(1-\M_{\rm sh})$ if the 
shock is weak. For a strong shock, $\qfq$ is a 
decreasing function of $\gamma$ ($\qfq\sim 1.7$ for $\gamma=5/3$):
\begin{equation}
\qfq =\left({2\gamma\over\gamma-1}\right)^{1\over4}
{2^{3\over2}\over 2^{1\over2}+\gamma^{1\over 2}(\gamma-1)^{1\over2}}.
\end{equation}
${\cal Q}$ is rewritten using the refraction coefficient $\rfr$:
\begin{eqnarray}
{\cal Q}&=& {\cs\over c_{\rm sh}}
 \rfr\left({1-\rfr\over1+\rfr}\right)^{1\over2}
\left({\cs^2-c_a^2\over\cs^2-c_av_a}\right)\qfq.
\end{eqnarray}
Since ${\cal Q}_{a}$ depends strongly on the frequency, the 
efficiency ${\cal Q}$ is a function of $\omega$. 
Schematically, $\rfr\sim 1$ for $\omega\ll\omega_{\rm cut}$, and 
$\rfr\sim 0$ for $\omega\gg\omega_{\rm cut}$, so that ${\cal Q}$ 
reaches its maximum value for a frequency close to the cut-off 
frequency. In particular ${\cal Q}_{\rm max}\gg 1$ if $\cs\gg c_{\rm sh}$:
\begin{equation}
\qfq{\cs\over c_{\rm sh}}>{\cal Q}_{\rm max}> 0.3 
\qfq\left(1 + {{\dot\M}_{\rm son}\cs
\over2{\dot\cs}}\right)^{-1} {\cs\over c_{\rm sh}}.\label{Qmax}
\end{equation}

\section{Contribution of the acoustic cycle\label{Acontrib}}

The complex frequency $\omega$ is decomposed into its real part 
$\omega_r$ and imaginary part $\omega_i$. Let us introduce the 
parameter $x>0$ and transform the eigenmodes equation as follows:
\begin{eqnarray}
\omega_i&\equiv&{1\over\tau_{EA}}\log|{\cal Q}x|,\label{defx}\\
{{\cal Q}\over|{\cal Q}|}
\e^{i\omega_r\tau_{EA}}&+&x^{1-{\tau_{AA}\over\tau_{EA}}}
{{\cal R}\over|{\cal Q}|^{\tau_{AA}\over\tau_{EA}}}
\e^{i\omega_r\tau_{AA}} =x.\label{xor}
\end{eqnarray}
The solutions of Eq.~(\ref{xor}) are bounded by the interval 
$x_{-}<x<x_{+}$, where $x_{-}<1<x_{+}$ are solutions of the equation:
\begin{equation}
1\pm x_{\pm}^{1-{\tau_{AA}\over\tau_{EA}}}
{|{\cal R}|\over|{\cal Q}|^{\tau_{AA}\over\tau_{EA}}} =x_{\pm},\label{xpm}
\end{equation}
from which we obtain Eq.~(\ref{wminmax}) satisfied by the 
corresponding growth rates $\omega_i^\pm$.
We deduce from Eqs.~(\ref{defx}) and (\ref{xpm}) that
\begin{equation}
\omega_i^\pm(|{\cal Q}|=1\mp|{\cal R}|)=0.
\end{equation}
The real part $\omega_r$ must satisfy the following phase equation, 
deduced from  Eq.~(\ref{xor}):
\begin{eqnarray}
\left\lbrack {\cal Q}{\sin\omega_r(\tau_{AA}-\tau_{EA})
\over \sin\omega_r\tau_{AA}}
\right\rbrack^{\tau_{AA}\over\tau_{EA}}=
{\cal R}{\sin\omega_r(\tau_{EA}-\tau_{AA})
\over \sin\omega_r\tau_{EA}},
\label{phase}\\
x = {{\cal Q}\over|{\cal Q}|}{\sin\omega_r(\tau_{AA}-\tau_{EA})
\over\sin\omega_r\tau_{AA}}.
\end{eqnarray} 
If the parameter $|{\cal R}|/|{\cal Q}|^{\tau_{AA}/\tau_{EA}}\ll1$, we 
can estimate the first order contribution of the purely acoustic 
cycle to the entropic-acoustic instability:
\begin{equation}
x-1\sim {{\cal R}\over|{\cal Q}|^{\tau_{AA}\over\tau_{EA}}}
\cos\omega_r\tau_{AA}.
\end{equation}
Together with Eq.~(\ref{defx}-\ref{xor}), we obtain
Eqs.~(\ref{omegarn}-\ref{omegain}).

\end{document}